\documentclass[sn-mathphys-num]{sn-jnl}

\usepackage{graphicx}%
\usepackage{multirow}%
\usepackage{amsmath,amssymb,amsfonts}%
\usepackage{amsthm}%
\usepackage{mathrsfs}%
\usepackage[title]{appendix}%
\usepackage{xcolor}%
\usepackage{textcomp}%
\usepackage{manyfoot}%
\usepackage{booktabs}%
\usepackage{algorithm}%
\usepackage{algorithmicx}%
\usepackage{algpseudocode}%
\usepackage{listings}%

\theoremstyle{thmstyleone}%
\theoremstyle{thmstyletwo}%

\theoremstyle{thmstylethree}%

\makeatletter
\newenvironment{breakablealgorithm}
  {
   \begin{center}
     \refstepcounter{algorithm}
     \hrule height.8pt depth0pt \kern2pt
     \renewcommand{\caption}[2][\relax]{
       {\raggedright\textbf{\ALG@name~\thealgorithm} ##2\par}%
       \ifx\relax##1\relax 
         \addcontentsline{loa}{algorithm}{\protect\numberline{\thealgorithm}##2}%
       \else 
         \addcontentsline{loa}{algorithm}{\protect\numberline{\thealgorithm}##1}%
       \fi
       \kern2pt\hrule\kern2pt
     }
  }{
     \kern2pt\hrule\relax
   \end{center}
  }
\makeatother

\begin{document}

\title[Article Title]{An Affine Equivalence Algorithm for S-boxes based on Matrix Invariants}


\author[]{\fnm{Xincheng} \sur{Hu}}\email{202221081008@std.uestc.edu.cn}

\author[]{\fnm{Xiao} \sur{Zeng}}\email{xiaozeng@uestc.edu.cn}

\author[]{\fnm{Zhaoqiang} \sur{Liu}}\email{zqliu12@gmail.com}

\author[*]{\fnm{Guowu} \sur{Yang}}\email{guowu@uestc.edu.cn}

\affil[]{\orgdiv{School of Computer Science and Engineering}, \orgname{University of Electronic Science and Technology of China}, \orgaddress{\city{Chengdu}, \postcode{611731}, \country{P.R.China}}}


\abstract{
\qquad We investigate the affine equivalence (AE) problem of S-boxes. Given two S-boxes denoted as $S_1$ and $S_2$, we aim to seek two invertible AE transformations $A,B$ such that $S_1\circ A = B\circ S_2$ holds. Due to important applications in the analysis and design of block ciphers, the investigation of AE algorithms has performed growing significance.

\qquad In this paper, we propose zeroization on S-box firstly, and the AE problem can be transformed into $2^n$ linear equivalence problems by this zeroization operation. Secondly, we propose standard orthogonal spatial matrix (SOSM), and the rank of the SOSM is invariant under AE transformations. Finally, based on the zeroization operation and the SOSM method, we propose a depth first search (DFS) method for
determining AE of S-boxes, named the AE\_SOSM\_DFS algorithm. Using this matrix invariant, we optimize the temporal complexity of the algorithm to approximately $\frac{1}{2^n}$ of the complexity without SOSM. Specifically, the complexity of our algorithm is $O(2^{3n})$. In addition, we also conducted experiments with non-invertible S-boxes, and the performance is similar to that of invertible S-boxes. Moreover, our proposed algorithm can effectively handle S-boxes with low algebraic degree or certain popular S-boxes such as namely AES and ARIA\_s2, which are difficult to be handled by the algorithm proposed by Dinur (2018). Using our algorithm, it only takes 5.5 seconds to find out that the seven popular S-boxes namely AES, ARIA\_s2, Camellia, Chiasmus, DBlock, SEED\_S0, and SMS4 are affine equivalent and the AE transformations of these S-boxes are provided.}




\maketitle

\section{Introduction}\label{sec1}

\noindent S-boxes are indispensable tools in symmetric key algorithms, serving as a crucial component for substitution operations. The concept of S-boxes was initially proposed by Claude Shannon, who first introduced their use in replacing plaintext with ciphertext in a seminal paper published in 1949 \cite{shannon1949communication}. Proper design of S-boxes is critical in symmetric key algorithms, as the security of cryptographic algorithms depends on
the quality of S-boxes. The non-linear and differential properties of the S-box provide a defense against linear and differential cryptanalysis, and remain an active research field \cite{zhang2014highly, lidl1997finite, koblitz1994course}. S-boxes are widely deployed in numerous symmetric key encryption algorithms, including Advanced Encryption Standard (AES) \cite{daemen2002design}, Data Encryption Standard (DES) \cite{encryption1999data}, and Blowfish \cite{tiessen2015security, youssef2005affine}. Regarding the security of S-boxes and Boolean functions, numerous scholars have proposed various construction methods \cite{feng2023novel, hua2021design, gao2024monomial, zhang2019group}.

The affine equivalence (AE) problem of S-boxes \cite{leander2007classification, sakalli2006affine} has important applications in the design and analysis of S-boxes, involving the search for affine transformations between two S-boxes. AE problem has been used to evaluate the security of symmetric key algorithms, and has been widely studied to improve their security \cite{sasdrich2016affine}. Additionally, the problem has close relations to other types of equivalence, such as EA-equivalence \cite{li2012nonexistence}, and CCZ-equivalence \cite{leander2007classification}. Furthermore, the equivalence problem can be applied to classify S-boxes and solve the partition problem \cite{leander2007classification, perrin2019partitions, saarinen2012cryptographic}, both of which carry significant implications for cryptology. The differential \cite{biham2012differential} and linear \cite{matsui1994linear} properties of S-boxes are invariant in the AE transformation, and these properties can help determine the AE. Many researchers have investigated the AE problem of Boolean functions \cite{meng2007analysis, zeng2023generalized}.

In this paper, we aim to determine whether two given S-boxes are affine equivalent or not. Specifically, let $S_1$ and $S_2$ be two S-boxes mapping $n$ bits to $m$ bits, we aim to find two affine transformations $A$ and $B$ such that $S_1 \circ A=B \circ S_2$. If no such affine transformations exist, $S_1$ and $S_2$ are not affine equivalent.


During the early stages of research in this field, the algorithm proposed by \cite{biryukov2003toolbox} employed a probabilistic approach to find the solution, and this approach is related to ``to and fro" algorithm \cite{patarin1998improved}. The authors also evaluated the complexity of the linear equivalence (LE) problem to be $O(n^32^{n})$, which increased to $O(n^32^{2n})$ when the S-boxes satisfy $S(0) = 0$. To optimize the AE algorithm, the minimum lexicographic order was utilized, resulting in a complexity of $O(n^32^{2n})$. In subsequent research, the algorithm presented by \cite{dinur2018improved} utilized the Boolean functions of S-boxes to study the properties of affine equivalent S-boxes. This approach led to the identification of a solution based on rank tables, with a complexity of $O(n^32^{n})$. Regarding the recent research on the AE algorithm, Anne Canteaut et al. \cite{canteaut2022recovering} investigates the AE of quadratic functions and provides an analysis of its computational complexity. Specifically, the authors demonstrate that the total computational complexity of the algorithm is approximately $O(2^{2n}(n^2+m^2)^\omega)$, where $\omega$ is the exponent of matrix multiplication.


The algorithm introduced in \cite{dinur2018improved} delivers better performance than \cite{biryukov2003toolbox}, albeit with the caveat that it is only applicable to S-boxes whose algebraic degree \cite{carlet2021boolean} exceeds $n-2$. And even in cases where the algebraic degree exceeds $n-2$, the algorithm cannot guarantee the determination of the corresponding affine equivalent transformations under AE. Moreover, for some popular S-boxes that have higher security on the web, this algorithm fails to provide a feasible AE transformation. 

To address the issues of the above algorithms, we propose a novel deterministic algorithm that consistently provides the correct result, and can deal with S-boxes with degree less than $n-2$. Based on the collected experimental measurements, the complexity of our algorithm is approximately $O(2^{3n})$.
 
Our approach begins by transforming the AE problem to LE problems via the zeroization operation at first. Specifically, based on $2^n$ ways to transform an S-box into a zero-point S-box, we only need to consider $2^n$ LE problems. Then, we propose the standard orthogonal spatial matrix (SOSM) to describe the linear property of S-boxes, as the rank of the SOSM is an invariant for the AE transformation. The SOSM of an S-box can capture the linear property of the S-box, and this property is formally described and proven through the Theorem 1. Although the definition of HDIM in \cite{perrin2016algebraic} is similar to this one, it has not been utilized for the AE problem and is defined specifically for permutations. In contrast, our proposed definition allows for more flexibility in generating high-dimensional elements and is even applicable for non-invertible S-boxes. Based on the zeroization operation and SOSM method, we propose the AE\_SOSM\_DFS algorithm to determine the AE of S-boxes, which can also be applied to non-invertible S-boxes. It is worth noting that we do not apply Gaussian elimination to determine the invertibility of $A$ and $B$. Instead, we use an array to record the occurrence of values in these affine transforms. In this manner, we can judge the invertibility of $A$ and $B$ in real time, with a temporal complexity of $O(1)$. By doing so, we can significantly reduce the number of computational branches, as described above.

The rest of the paper is organized as follows. In Section 3, we describe the transformation process in detail. In Section 4, we introduce the orthogonal spatial matrix (OSM) method, and we use the SOSM to initialize the relationship of $A$ and $B$. In Section 5, we present a deterministic algorithm using the zeroization operation and SOSM method for the AE problem. In Section 6, We conducted experimental evaluations of our algorithm on popular S-boxes on the sage website, random S-boxes, and S-boxes with low algebraic degree, and compared its performance with algorithms in \cite{biryukov2003toolbox, dinur2018improved}. Through our experiments, we have identified that seven S-boxes, namely AES, ARIA\_s2 \cite{kwon2003new}, Camellia \cite{aoki2001camellia}, Chiasmus \cite{schejbal2013reverse}, DBlock \cite{wu2015dblock}, SEED\_S0 \cite{lee2005seed}, and SMS4 \cite{Ltd2006}, exhibit affine equivalence, and we provide the AE transformations in Appendix A.

\section{Preliminaries}\label{sec2}

\noindent Given a vector $x = (x^{(0)},x^{(1)},...,x^{(n-1)})^T \in GF(2)^n$, where $GF(2)$ represents the binary field, the S-box $S : GF(2)^n \rightarrow GF(2)^m$ can be employed to map $x$ into another vector $y = (y^{(0)},y^{(1)},...,y^{(m-1)})^T \in GF(2)^m$. Both $x$ and $y$ can be represented by $n$-bit and $m$-bit binary integers, respectively. The addition between two vectors in $GF(2)^n$ is performed by computing their bitwise exclusive or (XOR) value. Similarly, the addition operation between two variables is represented by the XOR value of them.

\hfill

\noindent \textbf{Truth table.} Let $S : GF(2)^n \rightarrow GF(2)^m$ denote an S-box. The S-box can be represented by a $2^n \times m$ truth table $T_S$, where each element in the $i$-th row and the $j$-th column of the table corresponds to the $j$-th bit of the output $S(i)$ in binary form. The truth table is expressed as $T_S(i,j) (i = 0,1,...,2^n-1, j = 0,1,...,m - 1)$.






\hfill

\noindent \textbf{Input negation and output negation.}
Given an S-box $S : GF(2)^n \rightarrow GF(2)^m$. The input negation operation can transform $S(x)$ into $S'(x) = S(x + a)$, while the output negation operation can transform $S(x)$ into $S'(x) = S(x) + b$, where $a \in GF(2)^n$ and $b \in GF(2)^m$ are constant vectors.

\hfill

\noindent \textbf{Invertible affine transformation and AE of S-boxes.} Let $x \in GF(2)^n$ be a vector and $A : GF(2)^n \rightarrow GF(2)^n$ be a transformation. If $A$ can be represented as $A(x) = L(x) + c$, where $L$ is an invertible linear transformation represented by an $n \times n$ invertible matrix over $GF(2)$ and $c$ is a constant vector, then $A$ is an invertible affine transformation. Given two S-boxes $S_1$ and $S_2$, if there exist two invertible affine transformations $A$ and $B$ such that $S_1 \circ A = B \circ S_2$, then $S_1$ and $S_2$ are said to be affine equivalent.

\section{Zeroization operation} \label{sec3}

\noindent In this section, we introduce the zeroization method, which allows us to convert the AE problem into linear equivalence problems, leading to a simplification of the problem.

\hfill

\noindent \textbf{Zeroization operation of the S-box.} An S-box $S$ is a zero-point S-box if $S(0) = 0$. To determine whether two S-boxes $S_1$ and $S_2$ are affine equivalent, we seek two invertible affine mappings $L_1 \in GF(2)^{n \times n}, L_2 \in GF(2)^{m \times m}$ and $c_1 \in GF(2)^n, c_2 \in GF(2)^m$ such that $S_1(L_1(x) + c_1) = L_2(S_2(x)) + c_2$. However, enumerating all possible values of $c_1$ and $c_2$ to judge the linear equivalence of the resulting S-boxes leads to a complexity of $2^{n+m}$ times the complexity of the LE problem, which is prohibitive for large $n$. To overcome this challenge, we introduce the zeroization operation, which transforms an S-box into a zero-point S-box by operating on $c_1$ and $c_2$. By enumerating all possible zeroization operations, we can transform one S-box into a zero-point S-box and compare it with the other S-box in its zero-point form. This reduces the complexity of the algorithm to $2^n$ times linear equivalence algorithm.

Using input negation and output negation operations, we can transform the equation $S_1(L_1(x) + c_1) = L_2(S_2(x)) + c_2$ into the following form:

\begin{equation}
S_1(L_1(x) + c_1) + S_1(0) = L_2(S_2(x)) + c_2 + S_1(0). \label{pythagorean}
\end{equation}

Then, by letting $x=L_1^{-1}(c_1)$, we know that $S_1(0) = L_2(S_2(L_1^{-1}(c_1))) + c_2$, and by using the output negation operation, we can denote $S_1(x) + S_1(0)$ as $S_1'(x)$. Thus, we obtain the following expression:

\begin{equation}
S_1'(L_1(x) + c_1) = L_2(S_2(x) + S_2(L_1^{-1}(c_1))).\label{pythagorean}
\end{equation}

In the above equation, the term $c_2$ has been removed, and further equation yields the following expression:

\begin{equation}
S_1'(L_1(x)) = L_2(S_2(x + L_1^{-1}(c_1)) + S_2(L_1^{-1}(c_1))). \label{pythagorean}
\end{equation}

Under the input and output negation, we also represent $S_2(x + L_1^{-1}(c_1)) + S_2(L_1^{-1}(c_1))$ as $S_2'(x)$. Thus, we have 

\begin{equation}
S_1'(L_1(x)) = L_2(S_2'(x)). \label{pythagorean}
\end{equation}

Based on this, we can determine that both $S_1'$ and $S_2'$ are zero-point S-boxes. By transforming the original S-boxes $S_1$ and $S_2$ into zero-point S-boxes $S_1'$ and $S_2'$, we have simplified the problem. To solve the AE problem, we can enumerate all possible values of $c_1 \in \{0,1\}^n$, and determine whether $S_1'$ and $S_2'$ are linearly equivalent. This approach reduces the problem to a linear equivalence problem with a cost of $2^n$ loops, which can be solved using a linear equivalence algorithm. We discuss the details of the linear equivalence algorithms in subsequent sections.

\section{Transformation of S-box based on SOSM} \label{sec4}

\noindent In this section, we develop the definition of orthogonal spatial matrix and focus on the SOSM and its associated properties. Subsequently, we apply SOSM to transform S-boxes into new ones that can be judged with more necessary constraints.

\hfill

\noindent \textbf{Definition 1. (Orthogonal spatial matrix)} Given an S-box mapping $S : GF(2)^n \rightarrow GF(2)^m$, a set of $k$ linearly independent vectors $G = \{x_0,x_1,...,x_{k-1} \in GF(2)^n\}$ can be selected $(0 < k \le n)$. Using this set, we can construct a new $m \times k$ matrix $M(S,G)$, where the $(i,j)$-th element of the matrix is given by $\sum_{y \in W^\perp(x_j)} T_S(y,i)$, i.e., $M(S,G)_{i,j} = \sum_{y \in W^\perp(x_j)} T_S(y,i)$. We refer to this matrix as a $k$-orthogonal spatial matrix ($k$-OSM). In the special case when $k = n$, the matrix is referred to as an orthogonal spatial matrix (OSM).

\hfill

The present definition is of a general nature, and as such, we proceed to introduce a particular instance of OSM, which is better suited to embody the linear property and establish the relationship between the original affine transformations $A$ and $B$.

\hfill

\noindent \textbf{Definition 2. (Standard orthogonal spatial matrix)}  Consider an S-box $S : GF(2)^n \rightarrow GF(2)^m$. Let $G = \{e_0, e_1, \ldots, e_{n-1}\}$ be a set of orthonormal basis vectors, where $e_i$ represents the $i$-th unit vector. Then, the OSM associated with $S$ and $G$ is the standard orthogonal spatial matrix (SOSM), denoted by $M(S)$.

\hfill

The SOSM is an AE invariant, and we use its property in our algorithm, which can be characterized by Theorem 1.

\hfill

\noindent \textbf{Theorem 1.} Let $S_1$ and $S_2$ be two S-boxes defined on $GF(2)^n$ and $GF(2)^m$, respectively, and for each column of $T_{S_1}$ and $T_{S_2}$, the number of 0s and 1s are both even. If there exist invertible affine transformation matrices $L_1 \in GF(2)^{n \times n}$, $L_2 \in GF(2)^{m \times m}$, $c_1 \in GF(2)^n$, and $c_2 \in GF(2)^m$ such that $S_1(L_1(x) + c_1) = L_2(S_2(x)) + c_2$ for all $x \in GF(2)^n$, then we have the relation $M(S_1) \cdot (L_1^T)^{-1} = L_2 \cdot M(S_2)$.

\hfill

\noindent\textbf{\textit{Remark 1.}} Theorem 1 is employed to solve the case in which the sequence comprising 0s and 1s is of even parity, which is different from the applicability of the properties of HDIM in \cite{perrin2016algebraic}. For any bijective S-box, the condition is certainly satisfied. Consequently, in subsequent sections, all S-boxes satisfy this constraint naturally.

\hfill

Before presenting the proof of Theorem 1, we provide some useful lemmas. 


\hfill

\noindent \textbf{Lemma 1.}  Let $S : GF(2)^n \rightarrow GF(2)^m$ be an S-box, and for each column of $T_S$, the number of 0s and 1s are both even. For any $a \in GF(2)^n$ and $b \in GF(2)^m$, a new S-box $S' : GF(2)^n \rightarrow GF(2)^m$ can be obtained by $S'(x) = S(x+a)+b$. Then, $M(S) = M(S')$.

\hfill

\noindent \textbf{Proof of Lemma 1.} To prove this lemma, we consider the following two cases, which are $S'(x) = S(x + a)$ and $S'(x) = S(x) + b$.

\hfill

\noindent Case 1. $S'(x) = S(x + a)$ 

Using Definition 2, we have $M(S)_{i,j} = \sum_{y \in W^\perp(e_j)} T_S(y,i)$. Since $S'(x) = S(x + a)$, we have $T_{S'}(y,i) = T_S(y + a,i)$. Then, $M(S')_{i,j} = \sum_{y \in W^\perp(e_j)} T_{S'}(y,i) = \sum_{y \in W^\perp(e_j)} T_S(y+a,i) = \sum_{y+a \in W^\perp(e_j)} T_S(y,i)$.
Thus, we only need to prove that $\sum_{y+a \in W^\perp(e_j)} T_S(y,i) = \sum_{y \in W^\perp(e_j)} T_S(y,i)$. To prove this proposition, we only consider the $j$-th bit of $a$. If the $j$-th bit of $a$ is 0, we have that $y$ and $y+a$ belong to $W^\perp(e_j)$ simultaneously or not at all, and then $\sum_{y+a \in W^\perp(e_j)} T_S(y,i) = \sum_{y \in W^\perp(e_j)} T_S(y,i)$. On the other hand, if the $j$-th bit of $a$ is 1, there is only one item between $y$ and $y+a$ belonging to $W^\perp(e_j)$, and then $\sum_{y+a \in W^\perp(e_j)} T_S(y,i) = \sum_{y \notin W^\perp(e_j)} T_S(y,i)$. Because there are even number of 0s and 1s in the $i$-th column of $T_S$, we have that $\sum_{y \notin W^\perp(e_j)} T_S(y,i) = \sum_{y \in W^\perp(e_j)} T_S(y,i)$. Then, $\sum_{y+a \in W^\perp(e_j)} T_S(y,i) = \sum_{y \in W^\perp(e_j)} T_S(y,i)$, and the lemma holds.

\hfill

\noindent Case 2. $S'(x) = S(x) + b$ 

In this case, we first decompose $b$ into a binary vector form: $(b^{(0)},b^{(1)},...,$ \ 
$b^{(m - 1)})^T$, and express $b$ as a linear combination of ${e_0, e_1, \ldots, e_{m-1}}$, i.e., $b = \sum_{i = 0}^{m-1}b^{(i)}\cdot e_i$. Then, we have $M(S')_{i,j} = \sum_{y \in W^\perp(e_j)} T_{S'}(y,i) = \sum_{y \in W^\perp(e_j)}$ \ $(T_{S}(y,i) + b^{(i)}) = \sum_{y \in W^\perp(e_j)} T_{S}(y,i) = M(S)_{i,j}$.

\hfill

By proving the lemma in above cases, we prove that $M(S) = M(S')$ under $S'(x) = S(x + a) + b$. \hfill$\blacksquare$

\hfill

\noindent \textbf{Lemma 2.} Consider two S-boxes $S_1, S_2 : GF(2)^n \rightarrow GF(2)^m$, and for each column of $T_{S_1}$ and $T_{S_2}$, the number of 0s and 1s are both even. If there exist invertible linear transformations $L_1 \in GF(2)^{n \times n}$ and $L_2 \in GF(2)^{m \times m}$ such that $S_1 \circ L_1 = L_2 \circ S_2$, then we have $M(S_1) \cdot (L_1^T)^{-1} = L_2 \cdot M(S_2)$.

\hfill

\noindent \textbf{Proof of Lemma 2.} Following a similar approach as in the previous proof, we only need to consider two cases: $S_1 \circ L_1 = S_2$ or $S_1 = L_2 \circ S_2$.

\hfill

\noindent Case 1. $S_1 \circ L_1 = S_2$

To begin with, in the first case, we need to show that $M(S_1) \cdot (L_1^T)^{-1} = M(S_2)$. We decompose $L_1$ into a product of elementary column transformation matrices. According to basic linear algebra, we can write $L_1$ as $E^{(1)} \cdot E^{(2)} \cdot \cdots \cdot E^{(p)}$, where each $E^{(i)}$ denotes an elementary transformation matrix. An elementary transformation matrix $E_{(i,j)}$ is a matrix that adds the $i$-th column to the $j$-th column.

It can be easily shown that $E_{(i,j)} = E_{(i,j)}^{-1}$, and we now explore what $M(S_1) \circ (L_1^T)^{-1}$ means. To this end, we first prove that $M(S_1 \circ E_{(i,j)}) = M(S_1) \cdot E_{(i,j)}^T$.

 According to Definition 2, we have that $M(S_1)_{k,l} = \sum_{y \in W^\perp(e_l)} T_{S_1}(y,k)$. We then decompose $y$ into its binary components $(y^{(0)},y^{(1)},\dots,y^{(n-1)})$. We observe that the action of $E{(i,j)}$ on $y$ results in a modification of the $i$-th component of $y$ by adding its $j$-th component, i.e., $E{(i,j)}\cdot y = (y^{(0)},y^{(1)},\dots,y^{(i)}+y^{(j)},\dots,y^{(j)},\dots,y^{(n-1)})$ assuming $i<j$.
To prove $M(S_1 \circ E_{(i,j)}) = M(S_1) \cdot E_{(i,j)}^T$, we divide this proposition into two cases.

First, for $l \ne i$, we need to prove $\sum_{E_{(i,j)}\cdot y \in W^\perp(e_l)} T_{S_1}(y,k) = \sum_{y \in W^\perp(e_l)}$ \ $T_{S_1}(y,k)$. We find this equation apparently holds because $E_{(i,j)}$ does not affect the $l$-th component of $y$ and thus $E_{(i,j)}\cdot y \in W^\perp(e_l)$ if and only if $y \in W^\perp(e_l)$. Second, for $l=i$, we need to prove $\sum_{E_{(i,j)}\cdot y \in W^\perp(e_l)} T_{S_1}(y,k) = \sum_{y \in W^\perp(e_l)} T_{S_1}(y,k) + \sum_{y \in W^\perp(e_j)} T_{S_1}(y,k)$. We consider that $E_{(i,j)}\cdot y \in W^\perp(e_l)$ if and only if $y^{(j)}=0$ and $y \in W^\perp(e_l)$ or $y^{(j)}=1$ and $y \in W^\perp(e_l) \cap W^\perp(e_j)$. Therefore, we have 

\begin{equation*}
  \begin{aligned}
\noindent \sum_{E_{(i,j)}\cdot y \in W^\perp(e_l)} T_{S_1}(y,k) &= \sum_{y \in W^\perp(e_l) \cap W^\perp(e_j)} T_{S_1}(y,k) + \sum_{y \notin W^\perp(e_l) \cup W^\perp(e_j)} T_{S_1}(y,k) \\
&= (\sum_{y \in W^\perp(e_l) \cap W^\perp(e_j)} T_{S_1}(y,k) + \sum_{y \in W^\perp(e_l) - W^\perp(e_j)} T_{S_1}(y,k))\\
&+(\sum_{y \notin W^\perp(e_l) \cup W^\perp(e_j)} T_{S_1}(y,k) + \sum_{y \in W^\perp(e_l) - W^\perp(e_j)} T_{S_1}(y,k)) \\
&= \sum_{y \in W^\perp(e_l)} T_{S_1}(y,k) + \sum_{y \notin W^\perp(e_j)} T_{S_1}(y,k). 
  \end{aligned}
\end{equation*}

\noindent Since each column of $T_{S_1}$ contains an even number of zeros and ones, we have $\sum_{y \notin W^\perp(e_j)} T_{S_1}(y,k) = \sum_{y \in W^\perp(e_j)} T_{S_1}(y,k)$, and thus $\sum_{E_{(i,j)}\cdot y \in W^\perp(e_l)} T_{S_1}(y,k) = \sum_{y \in W^\perp(e_l)} T_{S_1}(y,k) + \sum_{y \in W^\perp(e_j)} T_{S_1}(y,k)$.

We have established that $M(S_1 \circ E_{(i,j)}) = M(S_1) \cdot E_{(i,j)}^T$. Specifically, we have $M(S_2) = M(S_1 \circ L_1) = M(S_1 \circ E^{(1)} \circ E^{(2)} \circ ... \circ E^{(p)}) = M(S_1) \cdot (E^{(1)})^T \cdot (E^{(2)})^T \cdot ... \cdot (E^{(p)})^T = M(S_1) \cdot (E^{(p)} \cdot E^{(p-1)} \cdot E^{(1)})^T = M(S_1) \cdot (L_1^{-1})^T$.

\hfill

\noindent Case 2. $S_1 = L_2 \circ S_2$

Our objective is to show that $M(L_2 \circ S_2) = L_2 \cdot M(S_2)$, where $L_2$ is a product of elementary column transformation matrices $E^{(1)}, E^{(2)}, ..., E^{(p)}$, and $S_2$ is a Boolean function. We begin by proving that $M(E_{(i,j)} \circ S_2) = E_{(i,j)} \cdot M(S_2)$ for any elementary column transformation matrix $E_{(i,j)}$. By examining the truth table of $E_{(i,j)} \circ S_2$, we observe that $T_{E_{(i,j)} \circ S_2} = E_{(i,j)} \cdot T_{S_2}$, since the transformation only involves two swapped columns of the truth table.

Using the definition of the SOSM and the fact that $T_{E_{(i,j)} \circ S_2} = E_{(i,j)} \cdot T_{S_2}$, we can conclude that $M(E_{(i,j)} \circ S_2)$ is obtained by adding the $j$-th row of $M(S_2)$ to the $i$-th row of $M(S_2)$. Therefore, we have shown that $M(E_{(i,j)} \circ S_2) = E_{(i,j)} \cdot M(S_2)$, and then $M(L_2 \circ S_2) = M(E^{(1)} \circ E^{(2)} \circ ... \circ E^{(p)} \circ S_2) = E^{(1)} \cdot E^{(2)} \cdot ... \cdot E^{(p)} \cdot M(S_2) = L_2 \cdot M(S_2)$.

\hfill

With the combination of above cases, we prove that $M(S_1) \cdot (L_1^T)^{-1} = L_2 \cdot M(S_2)$. \hfill$\blacksquare$

\hfill

\noindent\textbf{Proof of Theorem 1.} By applying above Lemmas 1 and 2, we know that $M(S_1(L_1(x) + c_1) + c_2) = M(S_1) \cdot (L_1^T)^{-1}(x)$ and $M(L_2(S_2)(x)) = L_2 \cdot M(S_2)(x)$ for all $x \in GF(2)^n$. Therefore, we have $M(S_1) \cdot (L_1^T)^{-1}(x) = L_2 \cdot M(S_2)(x)$ for all $x \in GF(2)^n$. \hfill$\blacksquare$

\hfill

In conclusion, we have shown that if two S-boxes $S_1$ and $S_2$ are affine equivalent, then their SOSMs $M(S_1)$ and $M(S_2)$ are related by the equation $M(S_1) \cdot (L_1^T)^{-1}(x) = L_2 \cdot M(S_2)(x)$ for all $x \in GF(2)^n$. This result provides an important insight into the relationship between the linear properties of S-boxes that are related by affine transformations.

\hfill

\noindent\textbf{Transformation of S-boxes using SOSM.} Based on linear algebra, we can perform a series of row-column linear transformations on a given matrix $L$ to obtain a special form 
$\left(
\begin{matrix}
E_r & O \\
O & O 
\end{matrix}
\right)$, where $r$ is the rank of $L$ and $E_r$ is the $r \times r$ identity matrix.

According to Theorem 1, we have $Rank(M(S_1)) = Rank(M(S_2))$, and we can set $r$ to be the rank of $M(S_1)$. Furthermore, there exist invertible matrices $A_1,A_2 \in GF(2)^{n \times n}$ and $B_1,B_2 \in GF(2)^{m \times m}$ such that $B_1 \cdot M(S_1) \cdot A_1 = B_2 \cdot M(S_2) \cdot A_2 = \left(\begin{matrix}E_r & O \\ O & O\end{matrix}\right)$. The matrices $A_1,A_2,B_1,B_2$ can be computed efficiently using Gaussian elimination.

Next, we can construct two new S-boxes $\widetilde{S_1}$ and $\widetilde{S_2}$ as follows: $\widetilde{S_1} = B_1 \circ S_1 \circ (A_1^T)^{-1}$ and $\widetilde{S_2} = B_2 \circ S_2 \circ (A_2^T)^{-1}$. It can be shown that $\widetilde{S_1}$ and $\widetilde{S_2}$ are affine equivalent if and only if $S_1$ and $S_2$ are affine equivalent. Therefore, we only need to test for AE between $\widetilde{S_1}$ and $\widetilde{S_2}$.

\hfill 

Based on SOSM, we transform S-boxes $S_1,S_2$ into new ones $\widetilde{S_1},\widetilde{S_2}$. Next, if they satisfies that $\widetilde{S_1}(L_1(x) + c_1) = L_2(\widetilde{S_2}(x)) + c_2$, we explore the relation between $L_1$ and $L_2$.

To describe the relation of $L_1$ and $L_2$, we express $L_1$ and $L_2$ as blocked matrices:

\begin{center}
$L_1 = \left( \begin{matrix}
L^{(11)}_1 & L^{(12)}_1 \\
L^{(21)}_1 & L^{(22)}_1
\end{matrix}
\right),
L_2 = \left( \begin{matrix}
L^{(11)}_2 & L^{(12)}_2 \\
L^{(21)}_2 & L^{(22)}_2
\end{matrix}
\right)$,
\end{center}

\noindent where $L^{(11)}_1,L^{(11)}_2$ are $r \times r$ matrices and $r$ is the rank of the SOSMs of $S_1$ and $S_2$. Then, the dimensions of the other blocks are also determined accordingly.

\hfill

After obtaining $\widetilde{S_1},\widetilde{S_2}$, we proceed to deduce the relationship equations between the sub-block matrices of $L_1$ and $L_2$, and the following theorem can explain it.

\hfill

\noindent \textbf{Theorem 2.} Let $\widetilde{S_1}$ and $\widetilde{S_2}$ be two affine equivalent S-boxes $(GF(2)^n \rightarrow GF(2)^m)$ obtained by transformations based on SOSM, and suppose that there exist $L_1 \in GF(2)^{n \times n}$, $L_2 \in GF(2)^{m \times m}$, $c_1 \in GF(2)^n$, and $c_2 \in GF(2)^m$ such that $\widetilde{S_1}(L_1(x) + c_1) = L_2(\widetilde{S_2}(x)) + c_2$. Then, we have $((L_1^{(11)})^T)^{-1} = L_2^{(11)}$ and $L_1^{(21)} = L_2^{(21)} = O$.

\hfill 

\noindent\textbf{Proof of Theorem 2.} We have $M(\widetilde{S_1}(L_1(x) + c_1)) = M(L_2(\widetilde{S_2}(x)) + c_2)$, and by Theorem 1, we obtain $M(\widetilde{S_1}) \cdot (L_1^T)^{-1} = L_2 \cdot M(\widetilde{S_2})$. Since $M(\widetilde{S_1}) = M(\widetilde{S_2}) = \left(\begin{matrix}E_r & O \\ O & O \end{matrix}\right)$, we obtain the following equation:

\begin{equation}
    \left(\begin{matrix}E_r & O \\ O & O \\ \end{matrix}\right) = \left(\begin{matrix}L^{(11)}_2 & L^{(12)}_2 \\ L^{(21)}_2 & L^{(22)}_2 \\ \end{matrix}\right) \cdot \left(\begin{matrix}E_r & O \\ O & O \\ \end{matrix}\right) \cdot \left(\begin{matrix}(L^{(11)}_1)^T & (L^{(21)}_1)^T \\ (L^{(12)}_1)^T & (L^{(22)}_1)^T \\ \end{matrix}\right).\label{pythagorean}
\end{equation}

By solving this equation, we obtain the following expression:

\begin{equation}
    \begin{cases}
    L^{(11)}_2 \cdot (L^{(11)}_1)^T = E_r \\
    L^{(21)}_2 \cdot (L^{(11)}_1)^T = O \\
    L^{(11)}_2 \cdot (L^{(21)}_1)^T = O \\
    L^{(21)}_2 \cdot (L^{(21)}_1)^T = O
    \end{cases}.\label{pythagorean}
\end{equation}

It is evident that both $L^{(11)}_2$ and $(L^{(11)}_1)^T$ are invertible, implying that $L^{(21)}_1 = L^{(21)}_2 = O$.
\hfill $\blacksquare$

\section{The AE
algorithm based on SOSM}

\noindent In this section, we present our AE algorithm. Based on the transformations elaborated in Section 3 and 4, we have developed AE\_SOSM\_DFS algorithm that can search all the solution spaces and assure the result is absolutely correct, and this algorithm is far more efficient than the general DFS algorithm without SOSM.

Firstly, using the SOSM method and the zeroization operation, the problem is tranformed to some linear equivalence problems, and the algorithm for solving the AE problem is called AE\_SOSM\_DFS algorithm (algorithm \ref{alg1}).

\hfill

\begin{breakablealgorithm}
\flushleft \caption{AE\_SOSM\_DFS Algorithm} \label{alg1}
\begin{algorithmic}[1]

\State Calculate the SOSM of $S_1$ and $S_2$ $(M(S_1)$ and $M(S_2))$
\State Calculate the rank of $M(S_1)$ and $M(S_2)$, denoted by $r_1$ and $r_2$
\If{$r_1 \ne r_2$}
\State Reject the AE of $S_1$ and $S_2$
\EndIf
\State Transform $S_1$ and $S_2$ into new S-boxes $\widetilde{S_1}$ and $\widetilde{S_2}$ such that $M(\widetilde{S_1})=M(S_1),M(\widetilde{S_2})=M(S_2)$
\State Initialize a new S-box such that $S_1'(x) = \widetilde{S_1}(x) + \widetilde{S_1}(0)$
\State $Success \leftarrow$ False
\For{every vector $a \in GF(2)^n$}
\State Initialize a new S-box such that $S_2'(x) = \widetilde{S_2}(x + a) + \widetilde{S_2}(a)$

\State Execute the LE\_SOSM\_DFS algorithm to check the linear equivalence of $S_1'$ and $S_2'$
\If{$Success$ = True}
\State \textbf{break}
\EndIf
\EndFor
\If{$Success$ = True}
\State Accept the AE of $S_1$ and $S_2$ and return AE transformations
\Else
\State Reject the AE of $S_1$ and $S_2$
\EndIf

\end{algorithmic}
\end{breakablealgorithm}

\hfill

\hfill

We have successfully transformed the problem of AE into that of linear equivalence. In the subsequent step, we introduce the LE\_SOSM\_DFS algorithm to solve the linear equivalence problem.

 \hfill

 To be convenient to express the vector whose bits after the $r$-th bit are all 0s, we define a new definition as follows.

 \hfill

 \noindent\textbf{Definition 3. (Suffix function)} Given a vector $x \in GF(2)^n$, the suffix function $suf$ with respect to $r$ and $x$, is defined by
 \begin{center}
 $suf(x,r) = \begin{cases}
 \prod_{i=r}^{n-1}{(x^{(i)} + 1)}, & r\ne n \\
 1, & r=n
 \end{cases},$
 \end{center}

 \hfill

 \noindent where $x^{(i)}$ is the $i$-th component of $x$.

According to Theorem 2, there exist relations between input vectors of $L_1$ and $L_2$, and we use the following lemma to optimize our LE algorithm.

 \hfill

 \noindent \textbf{Lemma 3.} Suppose $S_1$ and $S_2$ are two zero-point S-boxes which are transformed using SOSM, and there exist affine transformations $L_1$ and $L_2$ such that $S_1 \circ L_1 = L_2 \circ S_2$. Let $x \in GF(2)^n$ and $y \in GF(2)^m$ be the input vectors of $L_1$ and $L_2$, respectively, and it satisfies that $suf(x,r)=1,suf(y,r)=1$. Then, we have
 \begin{equation}
 ((x)_r)^T \cdot (y)_r = ((L_1(x))_r)^T \cdot (L_2(y))_r, \label{pythagorean}
 \end{equation}
 
 \noindent where $(x)_r$ denotes the $r$-dimensional vector obtained by retaining the first $r$ bits of $x$.

 \hfill 

\noindent\textbf{Proof of Lemma 3.} It is possible to express $((L_1(x))_r)^T \cdot (L_2(y))_r$ as $((x)_r)^T \cdot ((L_1)_r)^T \cdot (L_2)_r \cdot (y)_r$, where $L_r$ refers to the matrix that retains only the first $r$ rows and the first $r$ columns of $L$. According to Theorem 2, we have $((L_1)_r)^T \cdot (L_2)_r = E_r$, and hence, equation (\ref{pythagorean}) is established. This attribute enables us to eliminate numerous branches in our algorithm.
 \hfill $\blacksquare$

 \hfill

 Using lemma 3, we can easily check whether $L_1$ and $L_2$ are satisfied with Theorem 2, because checking two vectors costs less than two matrices. 

 \hfill

Let us consider how to solve linear equivalence problems, and it is to find the relation $S_1 \circ L_1 = L_2 \circ S_2$, where $S_1$ and $S_2$ are zero-point S-boxes $(GF(2)^n \rightarrow GF(2)^m)$ that have been transformed using SOSM. Similar to the algorithm in \cite{biryukov2003toolbox}, we utilize $C_{L_1}, C_{L_2}$ to record checked vectors of $L_1,L_2$ respectively, and $N_{L_1}, N_{L_2}$ to record new vectors that are required to check. Subsequently, we utilize sets $V_{L_1}, V_{L_2}$ to record the values that have assigned and determine whether the value space has been utilized fully without Gaussian elimination. When we get the value of a new vector on the mapping $L_1$ or $L_2$, we can deduce additional information about these mappings. We continuously process the vectors in sets $N_{L_1}$ and $N_{L_2}$ until there is no point in $N_{L_1},N_{L_2}$. By this way, if we put a new point to $N_{L_1}$ or $N_{L_2}$ and then we can calculate the value of more points. The specific processing procedure is denoted by \textsc{Work}.

\hfill

\begin{breakablealgorithm}

\begin{algorithmic}[1]

\State \textbf{procedure} \textsc{Work}($L_1,L_2$)
\While{$N_{L_1} \ne \varnothing$ \textbf{or} $N_{L_2} \ne \varnothing$}
\While{$N_{L_1} \ne \varnothing$}
\State pick $x \in N_{L_1}$
\If{$x \notin C_{L_1}$}
\If{$suf(x,r)=1$}
\If{\textsc{Check}$_{L_2}(x)$ = False}
\State $Reject \leftarrow$ True
\EndIf
\EndIf
\For{$x' \in C_{L_1}$}
\If{($x + x') \in N_{L_1}$ \textbf{and} $L_1(x + x') \ne L_1(x) + L_1(x')$ \textbf{or} $(L_1(x) + L_1(x')) \in V_{L_1}$}
\State $Reject \leftarrow$ True
\EndIf
\State $L_1(x + x') \leftarrow L_1(x) + L_1(x')$, $V_{L_1} \leftarrow V_{L_1} \cup \{L_1(x + x')\}$
\If{$S_2(x + x') \notin C_{L_2}$ \textbf{and} $S_2(x + x') \notin N_{L_2}$}
\If{$S_1(L_1(x) + L_1(x')) \in V_{L_2}$}
\State $Reject \leftarrow$ True
\EndIf
\State $L_2(S_2(x + x')) \leftarrow S_1(L_1(x) + L_1(x'))$, $V_{L_2} \leftarrow V_{L_2} \cup \{S_1(L_1(x) + L_1(x'))\}$
\EndIf
\EndFor
\EndIf
\State $N_{L_1} \leftarrow N_{L_1} \setminus \{x\}$, $N_{L_2} \leftarrow S_2(x + C_{L_1}) \setminus C_{L_2}$, $C_{L_1} \leftarrow C_{L_1} \cup (x + C_{L_1})$
\EndWhile

\While{$N_{L_2} \ne \varnothing$}
\State pick $y \in N_{L_2}$
\If{$y \notin C_{L_2}$}
\If{$suf(y,r)=1$}
\If{\textsc{Check}$_{L_1}(y)$ = False}
\State $Reject \leftarrow$ True
\EndIf
\EndIf
\For{$y' \in C_{L_2}$}
\If{$(y + y') \in N_{L_2}$ \textbf{and} $L_2(y + y') \ne L_2(y) + L_2(y')$ \textbf{or} $(L_2(y) + L_2(y')) \in V_{L_2}$}
\State $Reject \leftarrow$ True
\EndIf
\State $L_2(y + y') \leftarrow L_2(y) + L_2(y')$, $V_{L_2} \leftarrow V_{L_2} \cup \{L_2(y + y')\}$
\If{$S_2^{-1}(y + y') \notin C_{L_1}$ \textbf{and} $S_2^{-1}(y + y') \notin N_{L_1}$}
\If{$S_1^{-1}(L_2(y) + L_2(y')) \in V_{L_1}$}
\State $Reject \leftarrow$ True
\EndIf
\State $L_1(S_2^{-1}(y + y')) \leftarrow S_1^{-1}(L_2(y) + L_2(y'))$, $V_{L_1} \leftarrow V_{L_1} \cup \{S_1^{-1}(L_2(y) + L_2(y'))\}$
\EndIf
\EndFor
\EndIf
\State $N_{L_2} \leftarrow N_{L_2} \setminus \{y\}$, $N_{L_1} \leftarrow S_2^{-1}(y + C_{L_2}) \setminus C_{L_1}$, $C_{L_2} \leftarrow C_{L_2} \cup (y + C_{L_2})$

\EndWhile
\If{$| C_{L_1} | = 2^n$ \textbf{and} $| C_{L_2} | = 2^m$}
\State $Success \leftarrow$ True
\State \textbf{break}
\EndIf

\EndWhile

\end{algorithmic}
\end{breakablealgorithm}

\hfill

\hfill

\noindent\textit{\textbf{Remark 2.}} It is imperative to note that if the flag $Reject$ becomes true, the
current judgment should be terminated, and due to the probability of non-invertibility of $S_1,S_2$, we consider preimages $S_1^{-1},S_2^{-1}$ as: $S_1^{-1}(x)=\{y \in GF(2)^n : x=S_1(y)\},S_2^{-1}(x)=\{y \in GF(2)^n : x=S_2(y)\}$, which may be empty or multiple elements in them. While assigning values to new vectors, we should verify the equation (\ref{pythagorean}). To achieve this, the \textsc{Check} function, as described in the procedure \textsc{Work}, is utilized. However, it is worth noting that the exhaustive verification of all points that characterize in another mapping may prove to be excessively time-consuming. To address this issue, we use a novel technique that optimizes this process.

 \hfill 

 \noindent \textbf{The X-OR basis technique.} This technique is highly valuable for our algorithm, and we explicate its main idea in the following straightforward manner.

Consider a given set of $x \in GF(2)^n$. Our objective is to obtain a set of linearly independent vectors. If we exhaustively explore each vector in the set and utilize a new set for recording already-encountered vectors and their linear expansions, followed by verifying whether a new vector has already appeared, the complexity of this method would be $O(2^n)$ in the worst case. In order to mitigate the complexity of seeking linearly independent vectors, the X-OR basis technique employs a vector consisting of $n$ vectors to record the current set of linearly independent vectors. When a new vector is ready to be inserted into the X-OR basis, we can employ the present vectors to assess whether the new vector is represented by them. If the new vector is not represented by them, we can insert it, and the complexity of insertion is optimized to $O(n)$.

 The INITIAL and INSERT operations of the X-OR basis are outlined.

 \hfill

\begin{breakablealgorithm}

\begin{algorithmic}[1]

\State \textbf{procedure} INITIAL{}
\For{$i$ in range $[0,n-1]$}
\State set $basis(i)$ to empty vector
\EndFor

\State

\State \textbf{procedure} INSERT($x$)
\For{$i$ in range $[0,n-1]$}
\If{the $i$-th bit of $x$ is 0}
\State \textbf{continue}
\EndIf
\If{$basis(i)$ is empty vector}
\State $basis(i) \leftarrow x$
\State \textbf{break}
\EndIf
\State $x \leftarrow x + basis(i)$
\EndFor
\end{algorithmic}
\end{breakablealgorithm}

\hfill

\hfill

With the aid of this technique, we can rapidly verify the equation in Lemma 3 while assigning values to $L_1$ or $L_2$. Specifically, we can verify whether $((x_1)_r)^T \cdot (x_2)_r = ((L_1(x_1))_r)^T \cdot (L_2(x_2))_r$ holds between the new vector $x_1$ of $L_1$ (or $L_2$) and the vectors $x_2$ in the X-OR basis of $L_2$ (or $L_1$). 

Taking the X-OR basis of $L_2$ as a case study, the procedure for this operation are in procedure \textsc{Check}.

\hfill

\begin{breakablealgorithm}
\begin{algorithmic}[1]

\State \textbf{procedure} \textsc{Check$_{L_2}$}($x_1$)

\For{$i$ in range $[0,r-1]$}
\If{$basis(i) \ne 0$}
\State $val_1 \leftarrow x_1 + basis(i)$, $val_2 \leftarrow L_1(x_1) + L_2(basis(i))$
\State $flag \leftarrow 0$
\While{$val_1 \ne 0$ \textbf{and} $val_2 \ne 0$}
\State $flag \leftarrow flag + val_1$ \textbf{mod} $2 + val_2$ \textbf{mod} 2
\State $val_1 \leftarrow val_1$ / 2, $val_2 \leftarrow val_2$ / 2
\EndWhile
\If{$flag \ne 0$}
\State \textbf{return} False
\EndIf
\EndIf
\EndFor
\State \textbf{return} True

\end{algorithmic}
\end{breakablealgorithm}

\hfill

\hfill

Constrained by the cardinality of the set of vectors, the temporal complexity of the algorithm is $O(r^2)$.

\noindent \textbf{The LE\_SOSM\_DFS algorithm.}
We initialize sets $C,N,V$ before every attempt. Once $N_{L_1}$ and $N_{L_2}$ are empty, we will traverse the value of a new minimum point of $L_1$ or $L_2$ (the priority of point vectors can be determined based on their binary representation values, where those with lower values are favored due to their higher likelihood of satisfying the condition $suf(x,r)=1$). Expanding upon the algorithm presented in reference 
 \cite{biryukov2003toolbox}, we see that the initial linear equivalence algorithm, which exhibits a temporal complexity of $O(n^32^{2n})$ for verifying the zero-point S-boxes. Upon iterating the equivalence algorithm for $2^n$ instances, the total temporal complexity of the AE algorithm amounts to $O(n^32^{3n})$. However, according to \cite{canteaut2022recovering}, the computational complexity can be multiplied
by a factor $2^n$ in worst cases. 

We utilize a Depth First Search (DFS) algorithm to search for the solution. The current traversal value, $v$, necessitating assignment is subject to a validity check, whereby values falling outside the permissible range are promptly rejected. We can use a tree structure to represent the search process, and the efficiency of this DFS algorithm depends on the size of this tree. 

\begin{figure}[htp]
    \centering
    \includegraphics[width=0.7\textwidth]{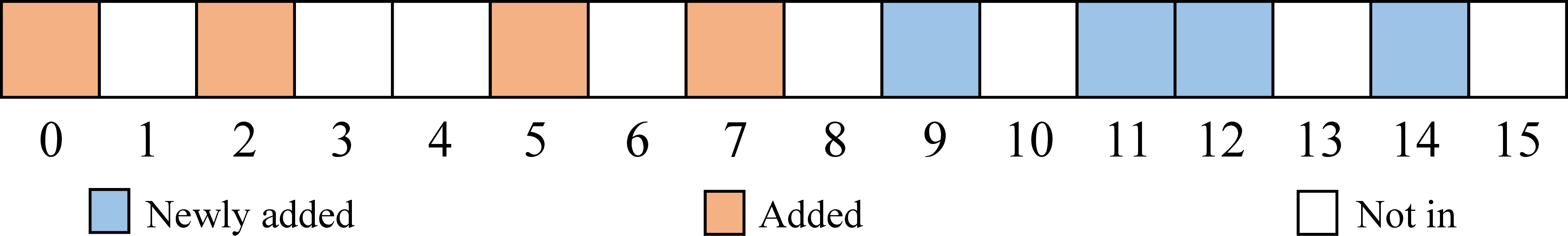}
    \caption{The difference of check set}
    \label{fig1}
\end{figure}

The use of SOSM accelerates the search process of above DFS algorithm, resulting in significant pruning of numerous branches. In light of the dependence between $L_1$ and $L_2$ for the first $r$ $(r>0)$ bits, the calculation complexity of the first $r$ bits can be seen as the optimization of SOSM. To be convenient to analyze the complexity, we assume that $n=m$. Additionally, according to reference \cite{kolchin1999random}, as $n$ approaches infinity, the probability that a random $n \times n$ Boolean matrix has rank $r$ is approximately given by the expression $\beta_r = 2^{-(r-n)^2} \cdot \alpha \cdot \Pi_{i=0}^{n-r}(1 - 1/2^i)^{-2}$, where $\alpha = \Pi_{i=1}^\infty(1 - 1/2^i) \approx 0.288$, and we know that the rank of SOSM is probably in proximity to $n$. Furthermore, we assume that $r=n$, and then we observe the variation in obtaining the value of a novel point vector. For example, when $n=4$, if we add a new point vector to the check set, the resulting distribution of the check set may be like Figure \ref{fig1}.

We denote the previous check set as $C_{L_1},C_{L_2}$. After adding a new point vector, $C_{L_1},C_{L_2}$ can be updated iteratively until no further vectors can be added, and we denote new check sets as $C'_{L_1},C'_{L_2}$. 
By each step of adding point vector, we can know that the quantity of new linearly independent equation (\ref{pythagorean}) is $\log_2|C'_{L_1}|\cdot\log_2|C'_{L_2}|-\log_2|C_{L_1}|\cdot\log_2|C_{L_2}|$. Based on the parity of equation (\ref{pythagorean}), we make the assumption that the probability of equation (\ref{pythagorean}) holding is 1/2. Then, the probability of accepting this step by SOSM method is approximately $\frac{1}{2^{\log_2|C'_{L_1}|\cdot\log_2|C'_{L_2}|-\log_2|C_{L_1}|\cdot\log_2|C_{L_2}|}}=\frac{2^{\log_2|C_{L_1}|\cdot\log_2|C_{L_2}|}}{2^{\log_2|C'_{L_1}|\cdot\log_2|C'_{L_2}|}}$, and the total probability of accepting from initial check set is $\frac{1}{2^{\log_2|C'_{L_1}|\cdot\log_2|C'_{L_2}|}}$ (multiply the probability of every step). In other words, the SOSM method can reduce the temporal complexity of the original DFS algorithm by a factor of $\frac{1}{2^{\alpha(n)}}$ in average, where $\alpha(n)$ denotes a funtion related to $n$ and has a value greater than 1. This function exhibits a tendency towards linearity with respect to $n$, and our observations from experiments of Section 6.2 indicate a complexity reduction by a factor of $\frac{1}{2^n}$ by using SOSM method.

  The process of LE\_SOSM\_DFS algorithm is presented in Algorithm \ref{alg2}.

\hfill

\renewcommand{\thealgorithm}{2}
\begin{breakablealgorithm}
\flushleft \caption{LE\_SOSM\_DFS Algorithm}
\label{alg2}

\flushleft{{\textbf{Input:}} S-boxes $S_1,S_2$

{\textbf{Output:}} If $S_1$ and $S_2$ are affine equivalent, accept them and give the transformation $L_1,L_2$. If not, reject it.}

\begin{algorithmic}[1]

\State $Success \leftarrow$ False, $Reject \leftarrow$ False
\State $L_1(0) \leftarrow 0, L_2(0) \leftarrow 0$, $C_{L_1} \leftarrow \{0\}, C_{L_2} \leftarrow \{0\}$, $V_{L_1} \leftarrow \{0\}, V_{L_2} \leftarrow \{0\}$, $N_{L_1} \leftarrow \varnothing, N_{L_2} \leftarrow \varnothing$
\State $v \leftarrow 0$
\While{$Success \ne$ True}
\If{$Reject =$ True}
\State Recover all the sets, X-OR basis and $v$ such that change in the last level
\State $v \leftarrow v + 1$(set $v$ to next value)
\If{$v = 2^n$}
\State go back to last \textbf{while} directly (e.q. \textbf{continue})
\EndIf
\State $Reject \leftarrow$ False
\Else
\State $v \leftarrow 0$
\EndIf
\State Pick a non-valued point $x$ with minimum lexical order of $L_1$
\If{$suf(x,r)=0$ \textbf{and} $suf(v,r)=0$ \textbf{or} $suf(x,r)=1$ \textbf{and} $suf(v,r)=1$ \textbf{and} \textsc{Check}$_{L_2}(x)$ = True}
\State $L_1(x) \leftarrow v$, $N_{L_1} \leftarrow N_{L_1} \cup x$, $V_{L_1} \leftarrow V_{L_1} \cup v$
\Else
\State Pick a non-valued point $y$ with minimum lexical order of $L_2$
\If{$suf(y,r)=0$ \textbf{and} $suf(v,r)=0$ \textbf{or} $suf(y,r)=1$ \textbf{and} $suf(v,r)=1$ \textbf{and} \textsc{Check}$_{L_1}(y)$ = True}
\State $L_2(y) \leftarrow v$, $N_{L_2} \leftarrow N_{L_2} \cup y$, $V_{L_2} \leftarrow V_{L_2} \cup v$
\Else
\State $Reject \leftarrow$ True
\EndIf
\EndIf
\State \textsc{Work}($L_1,L_2$)
\EndWhile
\If{$Success =$ True}
\State Accept the linear equivalence of $S_1$ and $S_2$
\State \textbf{return} $L_1$ and $L_2$
\Else
\State Reject the linear equivalence of $S_1$ and $S_2$
\EndIf
\end{algorithmic}
\end{breakablealgorithm}

\hfill

\hfill

\noindent \textbf{\textit{Remark 3.}} In order to prune as much as possible, we give priority to dealing with $x$ or $y$ that satisfy the $suf$ function equal to 1. Due to smaller degree of freedom of these points, we can get solution faster than seeking with no rules.

\section{Experiments and Results} \label{sec6_1}

\noindent Based on the Section 5, we have devised AE\_SOSM\_DFS algorithm based on SOSM for AE, and in this section, we examine its performance.

\hfill

\subsection{Applications on some of the popular S-boxes}

\noindent We implement the AE algorithms in \cite{biryukov2003toolbox} and \cite{dinur2018improved} respectively, and they are referred to in context as the ACAB and Dinur algorithms. The ACAB algorithm used here aims to find the S-box with the smallest lexicographical order among all the S-boxes that are linearly equivalent. Although they are applicable for determining random data, the Dinur algorithm is unable to handle some popular S-boxes or S-boxes with low algebraic degree, and we conduct experimental evaluations on popular S-boxes mainly.

We present the utilization of our algorithm to analyze some popular S-boxes from the sage website. Our algorithm and the ACAB algorithm can process these S-boxes, but the Dinur algorithm is ineffective in handling these particular S-boxes. 

This particular dataset on the website is frequently cited because its S-boxes exhibit favorable security properties. We inspected a total of 703 pairs of 8-bit S-boxes, comprising 38 distinct ones, including AES, Anubis, ARIA\_s2, BelT, Camellia, Chiasmus, CLEFIA\_S0, Crypton\_0\_5, Crypton\_1\_0\_S0, CS\_cipher, CSA, CSS, DBlock, E2, Enocoro, Fantomas, FLY, Fox, Iceberg, iScream, Kalyna\_pi0, Khazad, Kuznyechik, Lilliput-AE, MD2, newDES, Safer, Scream, SEED\_S0, SKINNY\_8, Skipjack, SNOW\_3G\_sq, SMS4, Turing, Twofish\_p0, Whirlpool, Zorro, ZUC\_S0. 

The AE\_SOSM\_DFS algorithm spend 4015 seconds to solve the AE problem of the 703 pairs of S-boxes, and 21 pairs of them exhibit AE. Nevertheless, in our experimental evaluations, the ACAB algorithm spent 15467 seconds finishing this task, and the Dinur algorithm is unable to derive a feasible solution. The using time and number of successful tests of these algorithms are presented in Table \ref{tab11}.

\begin{table}[h]
\centering
\caption{The using time and number of successful tests comparison of the AE\_SOSM\_DFS, ACAB and Dinur algorithms on 703 pairs of popular S-boxes}

\hfill

\begin{tabular}{c c c c}
\hline
&AE\_SOSM\_DFS&ACAB&Dinur\\\hline 
Time(s)&4015&15467&218\\
Number of successful tests&21&21&0\\
\hline
\end{tabular}
\label{tab11}
\end{table}

\noindent\textit{\textbf{Remark 4.}} Regarding the Dinur algorithm, we have observed that when the non-linearity of an S-box reaches a high level, its symbolic rank tends to become unitary. Taking AES as an example, when we set $d = n-2$, the symbolic rank of AES exhibits a specific pattern characterized by the occurrence of (8,7) tuples. As a result, in the subsequent flow of the Dinur algorithm, all of its high-support HSMs are identified within the same HG, rendering it impossible to establish a unique HSM. Because of this, the affine transformation cannot be uniquely determined. After setting $d$ to $n-1$, we observed that although the classification of symbolic ranks becomes diverse, it is still not possible to collect a sufficient number of unique HSMs to solve for the coefficients of affine equivalent transformations.

\hfill

We find out that seven of the S-boxes, specifically AES, ARIA\_s2, Camellia, Chiasmus, DBlock, SEED\_S0, and SMS4, are actually affine equivalent to each other, and the affine transformations are displayed in Appendix A.

\hfill

\subsection{Statistical analysis of randomized data}

\noindent The variable $Count$ is introduced in this study to represent the number of checkpoint operations executed by algorithm. As $Count$ increases, the traversal branches of the algorithms also increase proportionally, thereby allowing for the assessment of the effectiveness of different algorithms. In each case of $n$ $(4 \le n \le 10)$, 100 pairs of affine equivalent S-boxes are randomly generated and tested by AE\_SOSM\_DFS algorithm. 

Using above random data to test the complexity, we get log$_2(Count)$ and show them in Table \ref{tab2}, and the AE\_SOSM\_DFS algorithm guarantees the correctness of the obtained results. 

In addition, we will also investigate the impact on algorithm efficiency when removing the SOSM matrix optimization and zeroizations. The results are also shown in Table \ref{tab2}.

\hfill

\begin{table}[h]
\centering
\caption{The $Count$ (log$_2$) between the AE\_SOSM\_DFS algorithm and its counterparts without SOSM and zeroizations on random permutations}

\hfill

\begin{tabular}{c c c c}
\hline
$n$&AE\_SOSM\_DFS&Without SOSM&Without zeroizations\\\hline 
$4$&$9.15$&$14.02$&$11.89$\\
$5$&$12.34$&$17.60$&$15.21$\\
$6$&$15.25$&$21.73$&$18.08$\\
$7$&$18.48$&$25.39$&$21.51$\\
$8$&$21.37$&$28.59$&$24.31$\\
$9$&$24.08$&$31.77$&$27.30$\\
$10$&$27.57$&$35.34$&$30.40$\\
\hline
\end{tabular}
\label{tab2}
\end{table}

\begin{figure}[H]
    \centering
    \includegraphics[width=0.7\textwidth]{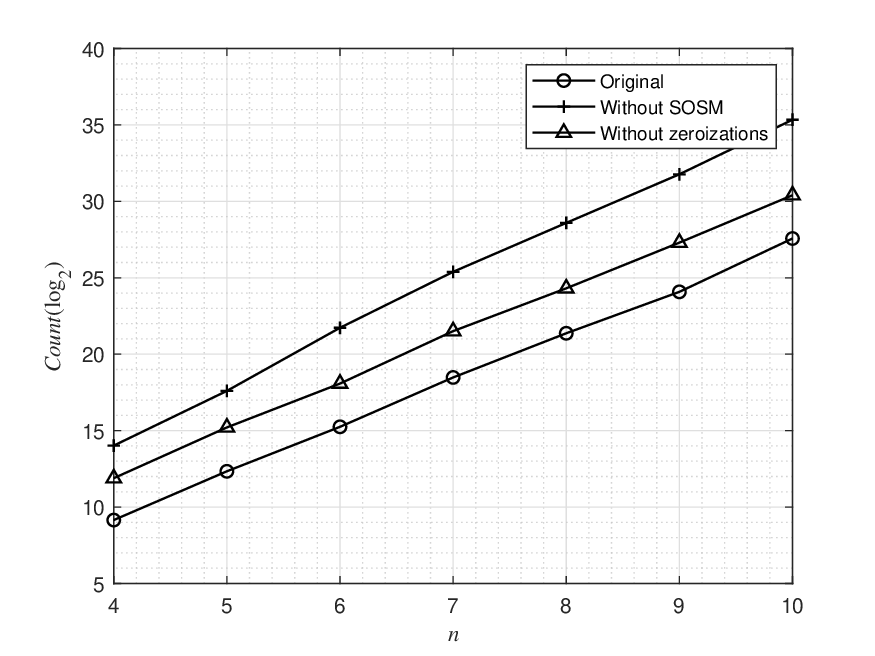}
    \caption{Complexity reduction from AE\_SOSM\_DFS to cases without SOSM and zeroizations}\label{fig2}
\end{figure}

\noindent\textit{\textbf{Remark 5.}} According to above experiences, we analyze the total complexity of AE\_SOSM\_DFS algorithm is approximately $O(2^{3n})$, and for random data, it accurately identifies the AE transformation. For random permutations, the Dinur algorithm exhibits higher efficiency compared to our algorithm, but the ACAB algorithm spend additional time finding the linear representation. Further detailed information is provided in Appendix B. By comparing these methods without SOSM or zeroizations, we have observed that the AE\_SOSM\_DFS algorithm is effective in reducing the number of branches. In the previous sections, we analyzed the total complexity of the AE\_SOSM\_DFS algorithm is approximately $\frac{1}{2^n}$ of the complexity of the algorithm without SOSM. The absence of zeroization does not impact the complexity of the algorithm, but it leads to a higher constant factor. The complexity is consistent with the findings displayed in Table \ref{tab2}. Upon reviewing Figure \ref{fig2}, it becomes apparent that the implementation of the SOSM and zeroization method expedites the process of solution searching. 

\hfill

\hfill

Additionally, our algorithm effectively addresses the AE problem for non-invertible S-boxes with an even distribution of 0s and 1s within each column of their respective truth tables. We develop a experiment on non-invertible S-boxes. Specifically, assuming that $n=m$, we generate 100 pairs of non-invertible S-boxes that satisfies above conditions, and for each input vector $x$ of the S-box $S$ we randomly generate a output vector $y \in GF(2)^n$. Then, we use AE\_SOSM\_DFS algorithm to determine their AE. The results of these experiments are presented in Table \ref{tab3}.

\hfill

\begin{table}[h]
\centering
\caption{The accuracy rate (AR) and average complexity ($Count$) for AE\_SOSM\_DFS algorithm on non-invertible S-boxes}

\hfill

\begin{tabular}{c c c}
\hline
$n$ & AR(\%) & Average $Count$\\\hline
$4$ & 100 & $279 \approx 2^{8.12}$\\
$5$ & 100 & $2835 \approx 2^{11.46}$\\
$6$ & 100 & $29083 \approx 2^{14.82}$\\
$7$ & 100 & $215443 \approx 2^{17.71}$\\
$8$ & 100 & $2550164 \approx 2^{21.28}$\\
$9$ & 100 & $26537542 \approx 2^{24.66}$\\
$10$ & 100 & $318024656 \approx 2^{28.24}$\\
\hline
\end{tabular}
\label{tab3}
\end{table}

\hfill

\noindent\textbf{\textit{Remark 6.}} Our experimental results show that the performance of AE\_SOSM\_DFS algorithm on non-invertible S-boxes is similar to the performance on invertible S-boxes, and we are able to provide the correct AE transformation for all the data.

\hfill

\subsection{Exploring the Algebraic Degree of S-Boxes}

\noindent Algebraic degree (AD) is a crucial indicator in evaluating the cryptographic strength of S-boxes. It measures the maximum degree of coordinate functions. In this study, we investigate the performance of our algorithm on S-boxes with different algebraic degrees.

The AD of certain extensively utilized S-boxes is as follows: 7 for AES \cite{do2019flexaead} and ZUC\_S1 \cite{zhou2011initialization}, 6 for CLEFIA\_S0 \cite{aoki2001camellia} and SKINNY\_8 \cite{datta2019breaking}, 5 for FLY \cite{karpman2016littlun} and ZUC\_S0 \cite{zhou2011initialization}, and 4 for CSS \cite{becker2004study}. To expand upon the analysis of AD's effect on our algorithm's performance, for each one of the above seven S-boxes, we randomly generate 100 S-boxes that are affine equivalent to it, and then use AE\_SOSM\_DFS to test the AE. The results are displayed in table \ref{tab4}.

\begin{table}[h]
\centering
\caption{The results of AE\_SOSM\_DFS on S-boxes with different ADs}

\hfill

\begin{tabular}{c c c}
\hline
Basic S-box & AR(\%) & Average $Count$\\\hline
AES (AD = 7) & 100 & $6975337 \approx 2^{22.73}$ \\
CLEFIA\_S0 (AD = 6) & 100 & $415633368 \approx 2^{28.63}$ \\
CSS (AD = 4) & 100 & $1898323\approx 2^{20.85}$ \\
FLY (AD = 5) & 100 & $360088920\approx2^{28.42}$ \\
SKINNY\_8 (AD = 6) & 100 & $229660269\approx 2^{27.77}$ \\
ZUC\_S0 (AD = 5) & 100 & $51865198\approx 2^{25.62}$ \\ 
ZUC\_S1 (AD = 7) & 100 & $6583764\approx 2^{22.65}$ \\\hline
\end{tabular}
\label{tab4}
\end{table}

\hfill 

We can see the performance of AE\_SOSM\_DFS algorithm may not be impacted much on S-boxes with low ADs, and the overall optimization performance remains notably effective. However, the Dinur algorithm is not capable of handling S-boxes with a low degree.

\section{Conclusions}

\noindent In this paper, we present an efficient algorithm for seeking AE between S-boxes. By selectively pruning branches, our algorithm achieves lower complexity than direct transformation checking. Further optimization opportunities may arise through the development of more effective pruning strategies. By using DFS algorithm with zeroization operation and SOSM method, our algorithm demonstrates a high degree of adaptability, thereby facilitating the fitting of S-boxes across a diverse range of scenarios.

\section*{Acknowledgment}

\noindent This work was supported by the National Natural Science Foundation of China under Grant 62172075.

\bibliography{sn-bibliography}

\appendix
\section{The affine transformation between some S-boxes.}

\noindent We determine the AE of AES, ARIA\_s2, Camellia, Chiasmus, DBlock, SEED\_S0 and SMS4, and it takes approximately 5.5 seconds on the laptop to check they are affine equivalent to each other and give the AE transformations. Upon substitution of $S_1$ with AES and $S_2$ with the other aforementioned S-boxes, the affine transformation coefficients in $S_1(L_1(x)+c_1) = L_2(S_2(x))+c_2$ are as follows.

  \hfill

  \noindent(1) AES ($S_1$) and ARIA\_s2 ($S_2$)

$(L_1 \mid c_1)=\left(
  \begin{array}{cccccccc|c}
    1 &0 &0 &0 &0 &0 &0 &0 &0\\
    0 &1 &0 &0 &0 &0 &0 &0 &0\\
    0 &0 &1 &0 &0 &0 &0 &0 &0\\
    0 &0 &0 &1 &0 &0 &0 &0 &0\\
    0 &0 &0 &0 &1 &0 &0 &0 &0\\
    0 &0 &0 &0 &0 &1 &0 &0 &0\\
    0 &0 &0 &0 &0 &0 &1 &0 &0\\
    0 &0 &0 &0 &0 &0 &0 &1 &0\\
  \end{array}
\right)$ \qquad
$(L_2 \mid c_2)=\left(
  \begin{array}{cccccccc|c}
  0 &1 &0 &1 &0 &0 &1 &0 &0\\
  1 &1 &0 &0 &1 &0 &1 &1 &0\\
  0 &0 &0 &1 &0 &0 &1 &0 &0\\
  1 &1 &0 &1 &1 &1 &0 &0 &0\\
  0 &0 &1 &0 &0 &0 &1 &0 &0\\
  0 &1 &0 &1 &1 &0 &0 &0 &1\\
  1 &1 &0 &1 &1 &1 &1 &0 &0\\
  1 &0 &0 &0 &1 &1 &0 &1 &0\\
  \end{array}
\right)$

\hfill

\noindent(2) AES ($S_1$) and Camellia ($S_2$)

  $(L_1 \mid c_1)=\left(
  \begin{array}{cccccccc|c}
  1 &0 &1 &1 &1 &1 &0 &0 &0\\
  1 &1 &0 &0 &0 &0 &0 &0 &0\\
  0 &1 &0 &1 &1 &1 &0 &0 &0\\
  1 &0 &1 &0 &0 &1 &1 &0 &0\\
  1 &0 &0 &0 &1 &0 &1 &0 &1\\
  0 &0 &0 &1 &0 &0 &0 &0 &0\\
  0 &0 &1 &1 &1 &0 &0 &0 &0\\
  1 &1 &1 &1 &1 &1 &1 &1 &0\\
  \end{array}
\right)$ \qquad
  $(L_2 \mid c_2)=\left(
  \begin{array}{cccccccc|c}
  1 &1 &1 &1 &1 &0 &0 &0 &1\\
  0 &1 &1 &0 &1 &0 &0 &0 &0\\
  1 &1 &1 &0 &0 &0 &0 &1 &1\\
  1 &1 &1 &1 &0 &1 &1 &1 &0\\
  1 &0 &1 &0 &0 &0 &0 &1 &1\\
  1 &1 &1 &0 &0 &1 &1 &0 &0\\
  1 &1 &1 &0 &0 &0 &1 &0 &0\\
  1 &1 &1 &0 &0 &1 &0 &0 &0\\
  \end{array}
\right)$

\hfill

\noindent(3) AES ($S_1$) and Chiasmus ($S_2$)

  $(L_1 \mid c_1)=\left(
  \begin{array}{cccccccc|c}
  1 &1 &0 &0 &1 &0 &0 &0 &0\\
  1 &1 &1 &0 &1 &0 &0 &0 &0\\
  1 &0 &1 &0 &0 &1 &1 &0 &1\\
  1 &0 &1 &0 &1 &1 &0 &0 &1\\
  0 &0 &1 &0 &1 &0 &0 &0 &1\\
  1 &1 &1 &1 &0 &0 &0 &0 &1\\
  0 &1 &1 &1 &0 &0 &0 &0 &0\\
  0 &1 &0 &0 &1 &0 &1 &1 &1\\
  \end{array}
\right)$ \qquad
  $(L_2 \mid c_2)=\left(
  \begin{array}{cccccccc|c}
  0 &0 &1 &0 &1 &0 &1 &1 &0\\
  1 &1 &1 &1 &1 &0 &1 &1 &1\\
  0 &1 &1 &1 &1 &1 &1 &0 &0\\
  0 &1 &1 &0 &0 &1 &0 &1 &0\\
  1 &1 &0 &0 &1 &1 &1 &1 &1\\
  0 &0 &1 &1 &0 &1 &1 &0 &1\\
  0 &1 &0 &1 &1 &0 &0 &0 &0\\
  1 &0 &0 &1 &1 &0 &1 &1 &0\\
  \end{array}
\right)$

\hfill

\noindent(4) AES ($S_1$) and DBlock ($S_2$)

  $(L_1 \mid c_1)=\left(
  \begin{array}{cccccccc|c}
  1 &0 &0 &0 &0 &0 &0 &0 &0\\
  1 &1 &0 &0 &0 &0 &0 &0 &0\\
  1 &0 &1 &0 &0 &0 &0 &0 &0\\
  1 &1 &1 &1 &0 &0 &0 &0 &0\\
  1 &0 &0 &0 &1 &0 &0 &0 &1\\
  1 &1 &0 &0 &1 &1 &0 &0 &0\\
  1 &0 &1 &0 &1 &0 &1 &0 &0\\
  1 &1 &1 &1 &1 &1 &1 &1 &0\\
  \end{array}
\right)$ \qquad
  $(L_2 \mid c_2)=\left(
  \begin{array}{cccccccc|c}
  0 &0 &0 &1 &1 &1 &0 &1 &0\\
  1 &0 &0 &1 &0 &0 &1 &1 &0\\
  0 &0 &1 &0 &1 &1 &0 &1 &0\\
  1 &1 &1 &1 &1 &1 &1 &1 &0\\
  0 &1 &1 &1 &0 &0 &0 &0 &0\\
  1 &0 &1 &1 &0 &1 &1 &1 &0\\
  1 &1 &0 &1 &0 &1 &1 &1 &1\\
  0 &0 &1 &1 &1 &0 &1 &1 &0\\
  \end{array}
\right)$

  \hfill

  \noindent(5) AES ($S_1$) and SEED\_S0 ($S_2$)

  $(L_1 \mid c_1)=\left(
  \begin{array}{cccccccc|c}
  0 &1 &1 &1 &0 &0 &0 &0 &0\\
  0 &0 &1 &1 &1 &1 &0 &0 &0\\
  0 &0 &0 &1 &1 &0 &0 &0 &0\\
  0 &0 &0 &1 &0 &1 &1 &0 &0\\
  0 &0 &1 &0 &1 &1 &1 &0 &0\\
  1 &0 &1 &1 &0 &0 &0 &0 &0\\
  1 &1 &0 &0 &1 &1 &0 &0 &0\\
  0 &0 &0 &0 &1 &0 &1 &1 &0\\
  \end{array}
\right)$ \qquad
  $(L_2 \mid c_2)=\left(
  \begin{array}{cccccccc|c}
  1 &0 &1 &0 &1 &0 &1 &0 &1\\
  1 &1 &0 &1 &1 &1 &0 &0 &1\\
  1 &0 &1 &0 &1 &0 &0 &1 &1\\
  1 &0 &0 &1 &0 &1 &1 &0 &1\\
  0 &0 &0 &1 &1 &0 &1 &1 &0\\
  1 &1 &1 &0 &0 &0 &1 &0 &0\\
  1 &1 &1 &0 &1 &1 &0 &1 &1\\
  1 &0 &1 &0 &0 &0 &1 &1 &0\\
  \end{array}
\right)$

\hfill

  \noindent(6) AES ($S_1$) and SMS4 ($S_2$)

  $(L_1 \mid c_1)=\left(
  \begin{array}{cccccccc|c}
  1 &1 &0 &0 &1 &0 &0 &0 &1\\
  1 &0 &0 &1 &1 &0 &0 &0 &1\\
  1 &0 &1 &1 &1 &0 &1 &0 &0\\
  1 &0 &1 &0 &0 &0 &0 &0 &1\\
  1 &0 &0 &0 &1 &1 &0 &0 &1\\
  0 &1 &1 &0 &0 &1 &0 &0 &1\\
  0 &1 &0 &0 &1 &1 &0 &0 &0\\
  1 &0 &0 &1 &0 &1 &0 &1 &1\\
  \end{array}
\right)$ \qquad
  $(L_2 \mid c_2)=\left(
  \begin{array}{cccccccc|c}
  1 &1 &0 &0 &0 &1 &0 &0 &0\\
  1 &0 &1 &0 &0 &0 &1 &0 &1\\
  0 &0 &0 &1 &1 &1 &1 &0 &1\\
  1 &0 &0 &0 &1 &0 &0 &0 &1\\
  1 &0 &0 &1 &0 &1 &1 &1 &0\\
  1 &1 &0 &1 &0 &0 &0 &0 &1\\
  0 &1 &1 &0 &1 &1 &1 &0 &1\\
  0 &1 &1 &0 &0 &0 &0 &0 &0\\
  \end{array}
\right)$

\hfill

\section{The possible limitation with the ACAB algorithm}

\noindent By randomly generating a permutation [2,7,3,4,1,0,6,5] with $n=3$, we can determine that its smallest LE representation is [1,0,2,3,4,6,5,7]. However, it may be a challenge for the ACAB algorithm to discover this LE representation by a small number of guesses \cite{canteaut2022recovering} due to the presence of numerous sections with variable values in sets $A$ and $B$. As a result, when executing the ACAB algorithm, it is possible to obtain alternative representations such as [1,0,2,4,3,5,6,7] or [1,0,2,3,4,6,5,7].

According to the ACAB algorithm, we perform the following assignments: $A(1) = 5$ and $B(1) = 2$. These assignments yield $S_R(1) = 0$ and $S_R(0) = 1$. In the subsequent step, if we assign $A(2) = 6$ and $B(2) = 6$, we obtain $S_R(2) = 2$, resulting in the correct linear representation. However, there is not a unique way to achieve $S_R(2) = 2$. For example, if we assign $A(2) = 1$ and $B(2) = 7$, we can still obtain that $S_R(2) = 2$. However, upon further analysis, we find that $A(3) = A(1) \oplus A(2) = 4$ and $B(3) = B(1) \oplus B(2) = 5$. Consequently, $S_R(3) = (B^{-1} \circ S \circ A)(3) = B^{-1}(1) \ne 3$, but based on the previously established correct linear representation, it can be inferred that $S_R(3) = 3$. Therefore, this alternative assignment does not correspond to the correct linear representation.

Therefore, when $n$ is large, obtaining the correct result solely through a small number of attempts becomes challenging. In order to ensure the identification of feasible solutions, it will take more time thorough exploration.

\end{document}